\documentclass{aa}  

\usepackage{graphicx}
\usepackage{txfonts}
\usepackage{siunitx,gensymb}
\usepackage{subcaption}

\usepackage{natbib,twoopt}
\usepackage[breaklinks=true,allcolors=blue]{hyperref} 
\bibpunct{(}{)}{;}{a}{}{,}             
\makeatletter
  \newcommandtwoopt{\citeads}[3][][]{\href{http://adsabs.harvard.edu/abs/#3}%
    {\def\hyper@linkstart##1##2{}%
     \let\hyper@linkend\@empty\citealp[#1][#2]{#3}}}
  \newcommandtwoopt{\citetads}[3][][]{\href{http://adsabs.harvard.edu/abs/#3}%
    {\def\hyper@linkstart##1##2{}%
     \let\hyper@linkend\@empty\citet[#1][#2]{#3}}}
  \newcommandtwoopt{\citepads}[3][][]{\href{http://adsabs.harvard.edu/abs/#3}%
    {\def\hyper@linkstart##1##2{}%
     \let\hyper@linkend\@empty\citep[#1][#2]{#3}}}
  \newcommandtwoopt{\citeyearads}[3][][]%
    {\href{http://adsabs.harvard.edu/abs/#3}
    {\def\hyper@linkstart##1##2{}%
     \let\hyper@linkend\@empty\citeyear[#1][#2]{#3}}}
\renewcommand*{\@fnsymbol}[1]{\ensuremath{\ifcase#1\or *\or \dagger\or \ddagger\or
   \mathsection\or \mathparagraph\or \|\or **\or \dagger\dagger
   \or \ddagger\ddagger \else\@ctrerr\fi}}
\makeatother
\usepackage{amstext}

\begin{document}

   \title{Interaction of a high-mass X-ray binary with the interstellar medium through stellar wind. The case of GX~301-2\thanks{Based on observations made with ESO Telescopes at the La Silla Paranal Observatory under program ID 099.D-0684(A)}\thanks{{\it Herschel} is an ESA space observatory with science instruments provided by European-led Principal Investigator consortia and with important participation from NASA.}}

   \author{J. Marchioro
          \inst{1}
          \and  S. Chaty\inst{1}
          \and A. Coleiro\inst{1}
          \and F. Fortin\inst{1}
          \and A. Simaz~Bunzel\thanks{Fellow of CONICET}\inst{2,3}
          }

   \institute{Université Paris Cité, CNRS, Astroparticule et Cosmologie, F-75013 Paris, France
        \and  Instituto Argentino de Radioastronom\'ia (CCT La Plata, CONICET; CICPBA; UNLP), C.C.5, (1894) Villa Elisa, Buenos Aires, Argentina
        \and Facultad de Ciencias Astron\'omicas y Geof\'{\i}sicas, Universidad Nacional de La Plata, Paseo del Bosque, B1900FWA La Plata, Argentina
             }
\titlerunning{}
 
  \abstract
   {GX~301-2 is a high-mass X-ray binary (HMXB) with strong stellar outflows. The evolution of these binaries can be closely related with the interstellar environment due to strong wind interactions.}
   {We try to constrain the energy injected in the interstellar medium by GX~301-2 through stellar wind using HAWK-I and \textit{Herschel} data.}
   {We analysed HAWK-I images in four different filters (Br$\gamma$, H$_2$, J, and Ks) and tried to retrieve signatures of the impact of GX~301-2 on its environment. We used \textit{Herschel} data to outline the interstellar medium and the Gaia DR3 catalogue to infer the proper motion of GX~301-2. Finally, we estimated the energy injected in the interstellar medium since the first supernova event of the HMXB.}
   {Using both HAWK-I and Herschel images, we deduce an approximation of the total mass injected from GX~301-2 in the interstellar medium of $M_{\rm inj}  = 3.05^{+0.05}_{-0.03} ~ 10^{-2} M_\odot$.}
   {}


   \maketitle
%

\section{Introduction}

Massive stars can create interstellar bubbles by sweeping material away \citep{vasquez_h_2005, petriella_molecular_2012}. They inject energy in the interstellar medium (ISM) through stellar winds at high terminal velocity, and can trigger stellar formation by generating overdensities in the ISM bubbles they created \citep{liu_triggered_2012}.

Studies show that 71\% of the stellar population in our galaxy spend some time of their evolution in a binary system \citep{sana_binary_2012}. High-mass X-ray binaries (HMXBs), which represent a step in the evolution of some massive binaries, are composed of a compact object (a neutron star or a black hole) orbiting a massive O/B star. A search for possible interaction with the ISM can give us a better understanding of the way in which HMXBs affect their surroundings by injecting energy in the ISM, which has already been proven for Cyg~X-1 \citep{gallo_dark_2005}.

GX~301-2 is a highly obscured HMXB composed of an accreting neutron star \citep{white_periodic_1976} orbiting a blue hypergiant of type B1Ia$^+$ \citep{kaper_vltuves_2006}. GX~301-2 is located at a distance of 3.6 $\pm$ 0.2 kpc, as inferred by \citet{bailer-jones_estimating_2021} using Gaia EDR3 astrometry. Studies from \citet{kaplan_long-wavelength_2006} and \citet{moon_rich_2007} showed that dust was present in the close vicinity of GX~301-2. In a subsequent study, \cite{servillat_herschel_2014} modeled a disk-like circumbinary environment encompassing the HMXB. \citet{huthoff_absence_2002} discovered an arm-like structure north of the source (referred to below as the arm) and hypothesized that it is in fact a bow shock coming from the interaction of GX~301-2 with its environment that sweeps the inner material toward this arm.
We here quantify the energy exchange between GX~301-2 as a donor and its environment. We search for infrared (IR) emission from gas and interstellar dust, and compute the energy budget of the interaction of the binary with the arm.

In Section 2 we review the photometric data we used in the study. Section 3 is an analysis of these observations, while in Section 4 we infer the mass transfer from the HMXB to the arm. These calculations are discussed and summarized in Section 5.

\section{Data acquisition}


\subsection{\textit{HAWK-I} data} 

Observations with the HAWK-I imager on the Very Large Telescope (VLT) \citep{pirard_hawk-i_2004} were performed between May 2-6, 2017, in four available energy bands, under ESO ID 099.D-0684(A) (PI Chaty). Photometric data have been collected using two broadband filters, J ($1.258\pm0.154$ \si{\micro\meter}), K$_s$ ($2.146\pm0.324$ \si{\micro\meter}) with a total exposure time of 1800~s and two narrowband filters, H$_2$ ($2.124\pm0.030$ \si{\micro\meter}) and Br$\gamma$ ($2.165\pm0.030$ \si{\micro\meter}), with a total exposure time of 2400~s. The data were reduced using the EsoReflex pipeline\footnote{\url{https://www.eso.org/sci/software/esoreflex/}} \citep{freudling_automated_2013} with the particular pipeline for HAWK-I imagery for dark subtraction and flat-field correction, as well as stacking images issued from jitter expositions.
The telescope was oriented so that GX~301-2 was located in the upper right quadrant of HAWK-I, in order to avoid a positioning at the edge between two quadrants. The produced images have a field of view of $7.5'\times 7.5'$. Some useful parameters of the data are summarized in Table \ref{tab:esoreflex}. 

\subsection{\textit{Herschel} data}

\textit{Herschel} \citep{pilbratt_herschel_2010} observations with the PACS \footnote{Photoconductor Array Camera and Spectrometer}\citep{poglitsch_photodetector_2010} instrument in imaging mode were carried out on August 2, 2011. Observations were made in the three available energy bands (blue 60-85 \si{\micro\meter}, green 85-130 $\si{\micro\meter},$ and red 130-210 \si{\micro\meter}) under Obs. ID 1342225124, 1342225125, 1342225126, and 1342225127 (PI Chaty). A summary of the observation properties is given in table \ref{tab:Herschel}. The data were reduced using the standard scripts dedicated to the reduction of PACS imaging mini-scan mode data, and implemented in the HIPE\footnote{\url{https://www.cosmos.esa.int/web/herschel/hipe-download\#documentation}} software.

\subsection{Spitzer data}

We retrieved archival data from the IRAC\footnote{Infrared Array Camera} instrument of the \textit{Spitzer} Space telescope \citep{werner_spitzer_2004,fazio_infrared_2004} at 4.5 $\si{\micro\meter}$ (CH2). This $1.3\degree\times44'$ image from the GLIMPSE survey \citep{benjamin_glimpse_2003} is centered on a region northwest of GX~301-2. More details about the parameters for the data acquisition can be found in \citet{churchwell_spitzer_2009}.



\begin{table*}[tb]
\centering
    \renewcommand{\arraystretch}{1.4} 

\begin{tabular}{lllll} \hline
    & H$_2$ & Br$\gamma$ & J & K$_s$ \\\hline
Spatial resolution$^1$ ($\arcsec$) & 0.61 & 0.72 & 1.00 & 0.92 \\
Saturation limit$^2$ (mag) & 10.85 & 10.89 & 8.67 & 6.60 \\
Total exposure time (s) & 2400 & 2400 & 1800 & 1800 \\
$N_{\rm frame}\times t_{\rm exp/frame}$ (s) & 30 $\times$ 80 & 30 $\times$ 80 & 30 $\times$ 60 & 30 $\times$ 60 \\
Airmass (start-end) & 1.423 -- 1.419 & 1.282 -- 1.283 & 1.496 -- 1.493 & 1.398 -- 1.395 \\
Jitter width (arcsec) & 20 & 20 & 20 & 20 \\
Seeing (corrected by airmass) & 0.78 & 1.08 & 1.01 & 1.07 \\\hline
\multicolumn{5}{l}{\small 1 Retrieved from HAWK-I instrument} \\
\multicolumn{5}{l}{\small 2 Derived from EsoReflex Exopsure Time Calculator} \\
\end{tabular}
\caption{Main parameters used for the data reduction in the four filters images for the HAWK-I instrument}
\label{tab:esoreflex}
\end{table*}

\begin{table*}[tb]
\centering
    \renewcommand{\arraystretch}{1.4} 

\begin{tabular}{lllll} \hline
    & 1342225124 & 1342225125 & 1342225126 & 1342225127 \\\hline
Mode & Scan map & Scan map & Scan map & Scan map \\
Duration (s) & 1350 & 1350 & 2705 & 2705 \\
Blue Channel & 70 $\mu$m band & 70 $\mu$m band & 100 $\mu$m band & 100 $\mu$m band \\\hline

\end{tabular}
\caption{Observational parameters for the Herschel/PACS images}
\label{tab:Herschel}
\end{table*}

\section{Data analysis}
The HAWK-I data were processed to detect features of interaction between GX~301-2 and the gas arm structure located north of the HMXB.
In the science images, we computed the instrumental point spread function (PSF) for each filter, and then subtracted it. Using PyRAF \footnote{\url{https://pyraf.readthedocs.io/en/latest/\#}}, we applied PSF subtraction to each detected star using the \textit{findstar} routine to detect stars above a given threshold, referenced in Table \ref{tab:PSF}, along with the number of stars that was subtracted for each filter. To remove the stars, we used the crowded field photometry method following \citet{massey_users_1992}. A selection of the brightest nonsaturated and sufficiently isolated stars was made by hand, and a first instrumental PSF was retrieved for the stars using the \textit{psf} routine. We removed the close neighbors of the bright stars by clustering regions using the \textit{group} command and subtracted the faintest stars grouped with the bright stars that were used for the first PSF using the \textit{substar} routine. This produced a cleaner second iteration of the instrumental PSF. It was then applied to subtract all the stars detected by the \textit{findstar} routine. The PSF subtraction did not work well on saturated stars, including GX~301-2, nor on stars close to the edge of quadrants. However, this did not affect the study because we focus on a faint diffuse region. The resulting images to which the subtracted PSF was applied enabled us to detect objects with a signal-to-noise-ratio (S/N) threshold 20 times lower than for the science images coming from the Esoreflex pipeline.
\medskip

\begin{figure*}
   \centering
   \begin{subfigure}[b]{0.3\linewidth}
    \includegraphics[width=\linewidth]{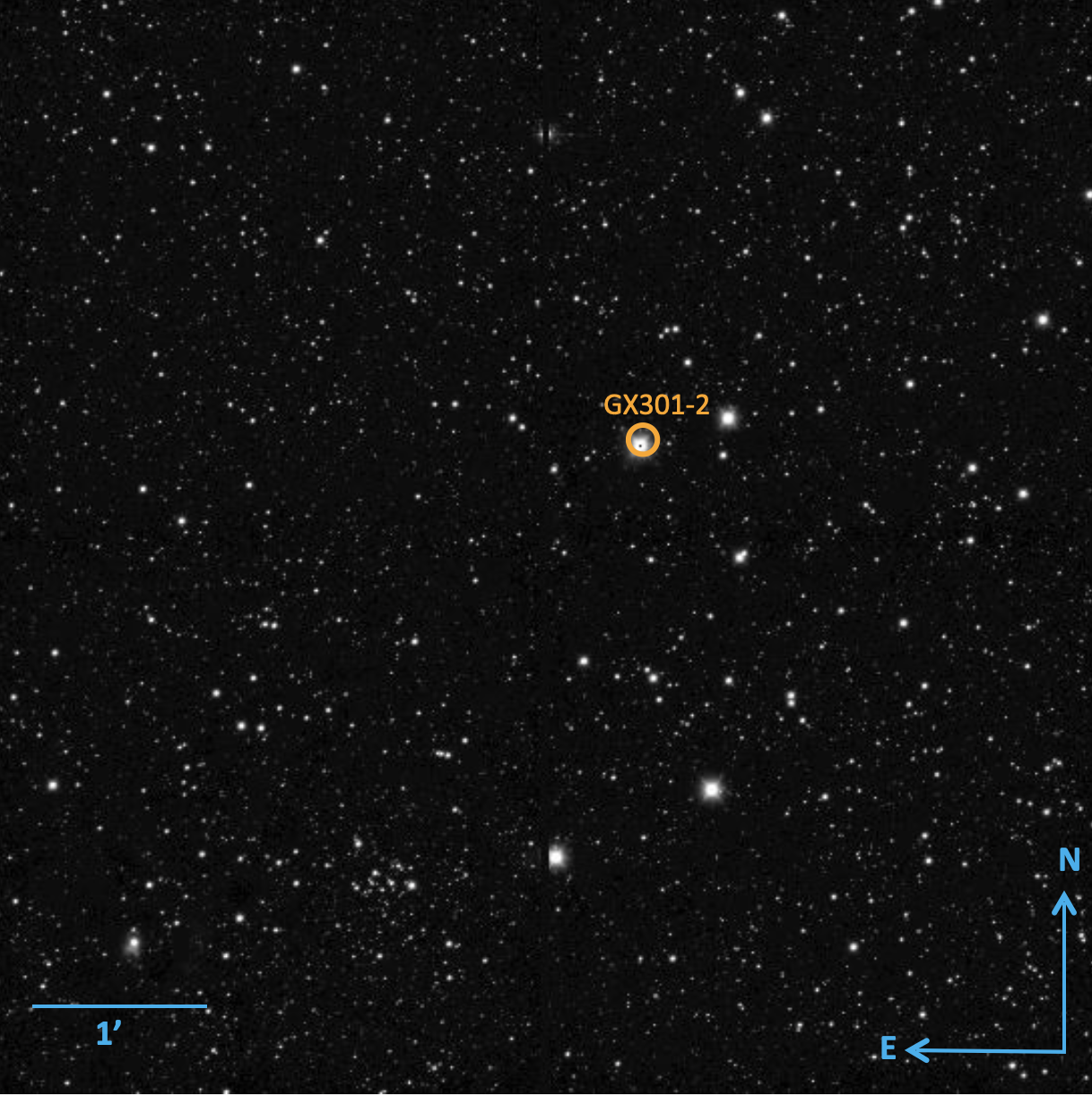}
   \end{subfigure}
   \hfill
   \begin{subfigure}[b]{0.3\linewidth}
    \includegraphics[width=\linewidth]{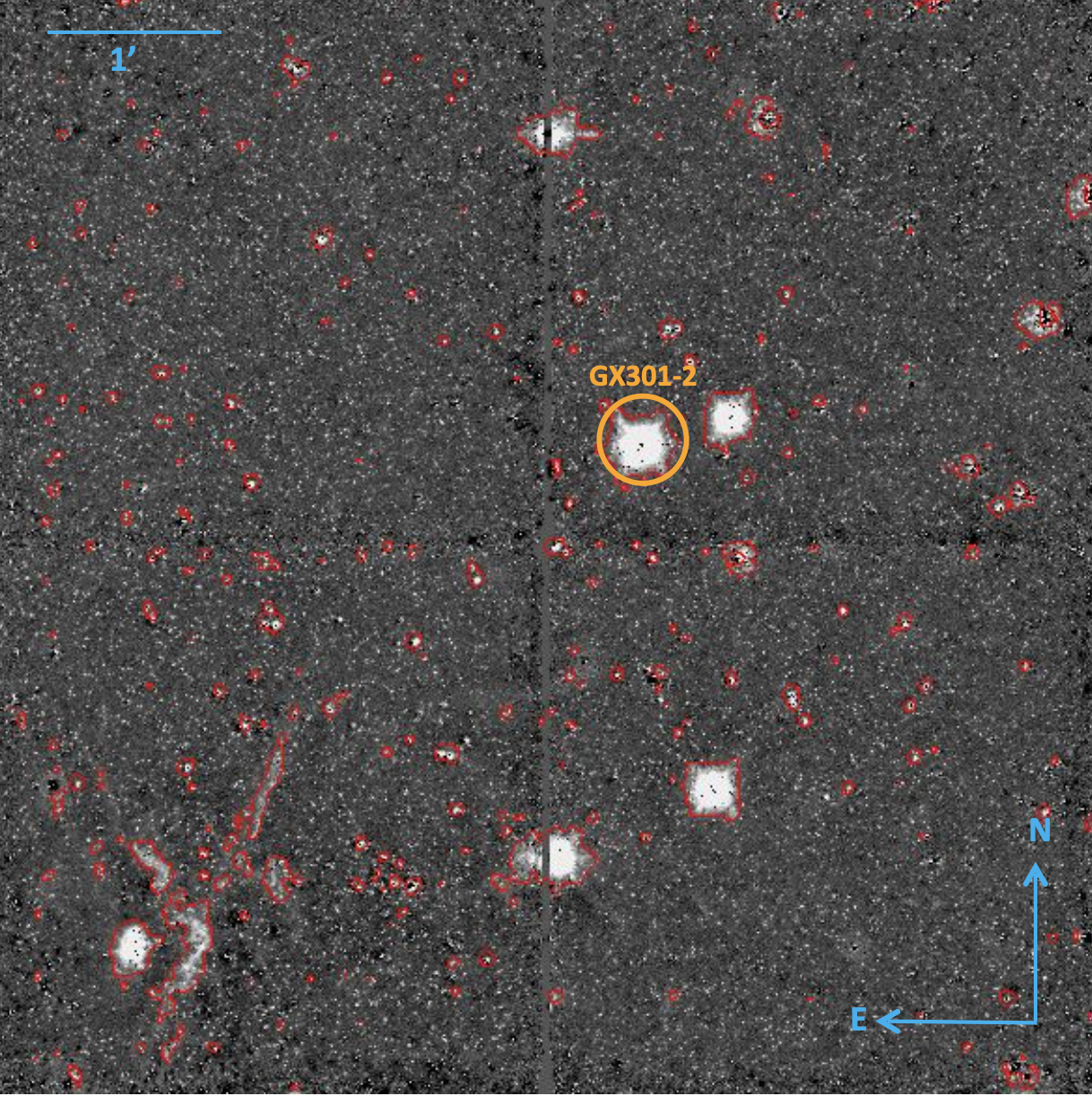}
   \end{subfigure}
   \hfill
   \begin{subfigure}[b]{0.3\linewidth}
    \includegraphics[width=\linewidth]{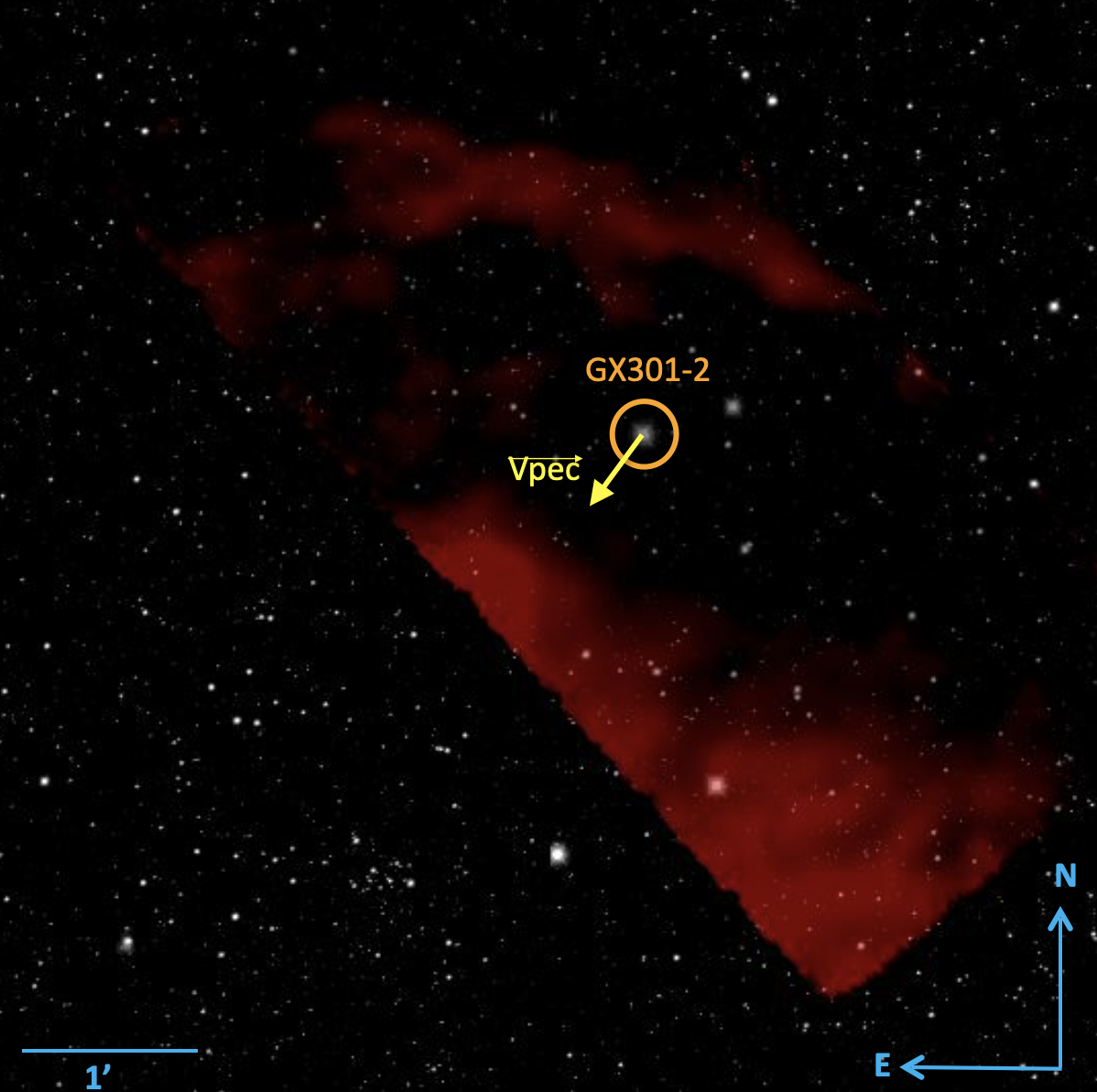}
   \end{subfigure}
   
   \caption{HAWK-I image in the H$_2$ filter. Panel a (left) is the raw scientific image.
   Panel b (middle) is the image after PSF subtraction, using contours to highlight brighter features. Panel c (right) is the raw scientific image, and the \textit{Herschel}/PACS image is overlaid. The yellow arrow indicates the peculiar velocity of GX~301-2 as retrieved from the Gaia EDR3 parallax study. GX~301-2 is marked in all three images. The contour corresponds to a 2.124 $\si{\micro\meter}$ surface brightness: 10 MJy/sr.}
   \label{fig:Hawki}%
\end{figure*}

\begin{figure*}
   \centering
   \begin{subfigure}[b]{0.45\linewidth}
    \includegraphics[width=\linewidth]{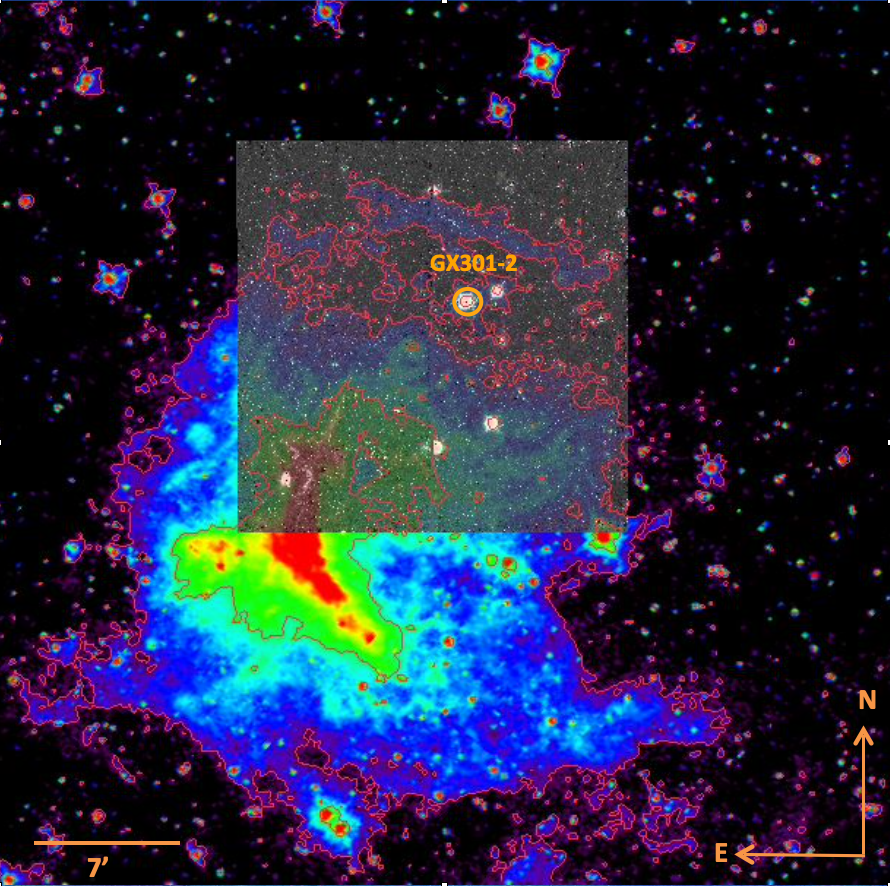}
   \end{subfigure}
   \hfill
   \begin{subfigure}[b]{0.45\linewidth}
    \includegraphics[width=\linewidth]{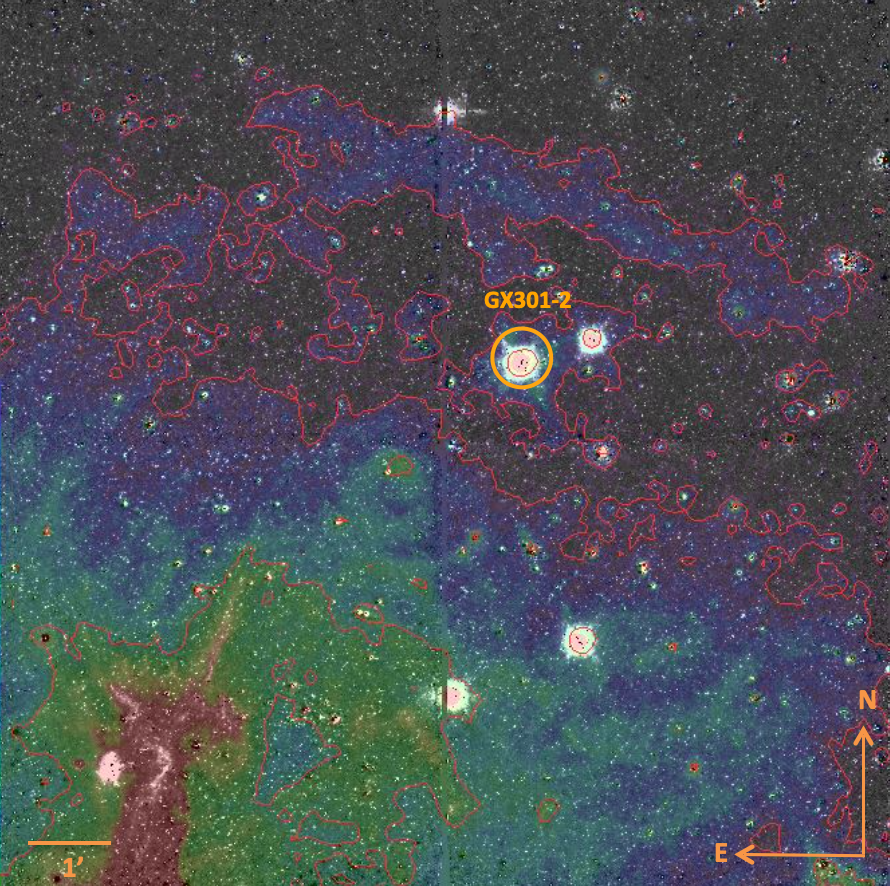}
   \end{subfigure}
   
   \caption{HAWK-I image in the H$_2$ filter. Panel a (left) is the subtracted image and the \textit{Spitzer} archive image in the background, which confirms at a larger scale that the bright residual in the bottom left quadrant is an astrophysical feature. We used contours to highlight the relevant arm-like structure that is present in the \textit{Spitzer} image, but missing in the HAWK-I image. Panel b (right) is a zoom centered on the field of view of HAWK-I. GX~301-2 is marked in the two images. Contours correspond to the 4.5 $\si{\micro\meter}$ surface brightness: 100 and 230 MJy/sr.}
   \label{fig:subki}%
\end{figure*}

Panels a and b of figure \ref{fig:Hawki} show the H$_2$ filter image before and after PSF subtraction. No clear evidence of the surrounding arm structure (which is shown in panel b of figure \ref{fig:subki} with \textit{Spitzer} data) was detected. However, a brighter area was detected in the bottom left quadrant, corresponding to a dust clump detected in the ATLASGAL catalogue \citep{urquhart_atlasgal_2018}, which is shown in \ref{fig:subkil}.

We used contours to search for gas structures in the HAWK-I images, with an average value higher than the background noise. We used $\frac{N_{\rm thresh}-N_{\rm BG}}{N_{\rm BG}} = 0.3\%$. This did not lead to any detection of the arm structure. We then used a 2D wavelet decomposition to perform a multiscale analysis of the image. The original image was passed through high and low filters in the two spatial directions at each iteration, so that for scale $n$, $4^n$ images were produced, and every iteration increased the scale that was probed. Our image had a 7 pc$^2$ field of view as given by the angular distance and the distance of GX~301-2 from \citet{bailer-jones_estimating_2021}, with a pixel size of $1.6 \times 10^{-3}$pc$^2$. Each scale probed an area twice as large as the previous scale, therefore 12 iterations to the wavelet decomposition process were possible. At order 10, we highlight structures of $\approx 1$ pc, but this requires creating $4^{10}$ images because they are passed through the 2D filters. However, the analysis of the decomposed images did not reveal the arm structure using this method. 

We overlaid the \textit{Herschel}/PACS data on the HAWK-I images, covering a field of view of $\approx 8.0\arcmin\times 3.5\arcmin$. The $Herschel$ image clearly displays the arm structure, so that every information concerning its geometry was retrieved using the data from \textit{Herschel}/PACS. However, due to the field of view, only a fraction of the arm structure was detected and used for the study. After characterizing and subtracting the PSF for this instrument, we applied a median filter to the residual frames  following \cite{povich_multiwavelength_2007}. After processing, the \textit{Herschel} image displayed the arm structure in the 70 to 160 $\si{\micro\meter}$ range. Figure \ref{fig:rawkel} shows the superimposition of HAWK-I H$_2$ filter and \textit{Herschel} $100\si{\micro\meter}$ images.

\begin{figure*}
   \centering
   \includegraphics[width=.8\linewidth]{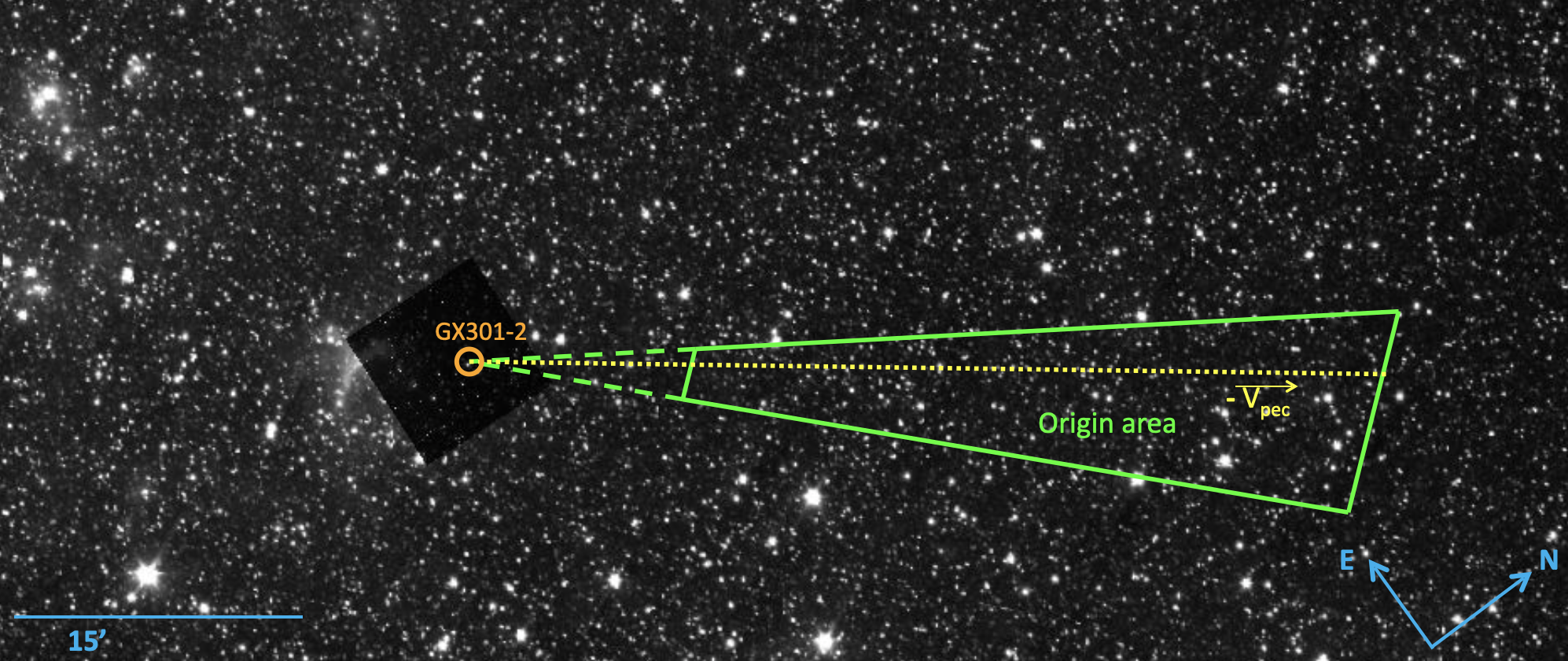}
   \caption{Spitzer image of a wider portion of the sky, containing the original area of GX~301-2. The green polygon is the uncertainty cone for GX~301-2 peculiar velocity, truncated orthogonally to $-\protect\overrightarrow{V_{\rm pec}}$ so that the supernova event occurred between~250~kyr~and~1~Myr. The yellow line is the average line for the trajectory. The HAWK-I image is to scale.}
   \label{fig:orig}%
\end{figure*}

\citet{fortin_constraints_2022} used Gaia EDR3 astrometric data to infer the peculiar velocity of \mbox{GX~301-2,} which corresponds to the systemic velocity corrected for Galactic rotation. While only the magnitude of the peculiar velocity vector is published, we asked the authors for the full vector and its 1\,$\sigma$ uncertainties (2022, priv. comm.), which is plotted in panel c of figure \ref{fig:Hawki}. Figure \ref{fig:orig} uses the vector to retrace the origin of GX~301-2, as well as its 1\,$\sigma$ uncertainties. Changes in the orbital parameters of the binary during a supernova explosion are associated with a momentum kick (natal kick) that is imparted to the newborn neutron star. The peculiar velocity we retrieved takes this natal kick into account and enables us to shape an origin area for the first supernova (SN) event. This leads to the conclusion that GX~301-2 originates far away from the gaseous region (see panel c of figure \ref{fig:Hawki}).

\begin{table}
\centering
\begin{tabular}{lllll} \hline
    & H$_2$ & Br$\gamma$ & J & K$_s$ \\\hline
$\sigma_\mathrm{sky}$ (ADU) & 9.39 & 16.66 & 32.99 & 100.72 \\
Threshold (in $\sigma_\mathrm{sky}$) & 18 & 4.5 & 1.4 & 13.5 \\
\# of stars & 12593 & 10520 & 2962 & 8891 \\\hline
\end{tabular}
\caption{Main parameters used for PSF subtraction of stars in the four filters images for the HAWK-I instrument.}
\label{tab:PSF}
\end{table}

\section{Estimation of the energy injected into the ISM}

We computed the energy budget of the stellar winds that are injected into the ISM. To do this, we first built a simple static model that we then improved using a dynamical evolution of the system.

To account for stellar winds coming from GX~301-2 and injected into the ISM, we assumed a constant wind flow rate, starting from the first SN event, which was estimated to occur around~$0.5^{+0.5}_{-0.25}$~Myr in the past (\citet{simaz_origin_2022}) and a spherically symmetric distribution of the wind.
We modeled the interaction as a continuously expanding bubble of matter crossing the ISM.
We only considered the arm-like gas structure that is susceptible to interacting with the stellar winds.

We approximated the arm-like structure as a cylinder with a symmetry axis coplanar to GX~301-2, to help us estimate the distances. We find a diameter of~0.5~pc and a height of~1.6~pc given the distance to GX~301-2 of 3.6~kpc. The cylinder was located at 1.3~pc from the HMXB, which was first considered as a fixed object, with a mass-loss rate $\dot{M}=10^{-5}$ M$_\odot$~yr$^{-1}$ driven by stellar winds with a terminal velocity of $v_{\rm w}=305$~km~s$^{-1}$, as described in \mbox{\citet{servillat_herschel_2014}.}

We estimated the time it takes for the stellar winds to reach the ISM to be 4~kyr, knowing the current distance between GX~301-2 and the arm gas structure. This value was compared to the time of integration, starting at the first SN event, and we decided to neglect the traveling time for the stellar winds because of their high velocity. The solid angle formed by the cylinder viewed from GX~301-2 is a rectangular shape of $\Omega = \frac{\mathnormal{a}}{l^2}= 0.47$~$\si{\steradian}$, with $\mathnormal{a}$ the effective area of the cylinder from GX~301-2 point of view, and $l$ is the distance to GX~301-2, assuming coplanarity with the arm.
We computed $l$ with the knowledge of the angular distance between GX~301-2 and the edge of the arm, as well as the distance of GX~301-2 from us, as retrieved by \citet{bailer-jones_estimating_2021}.

When we assume a uniform and isotropic ejection of the stellar wind from GX~301-2, only a fraction $x=\frac{\Omega}{4\pi}$ is perceived by the arm.
Thus, the total mass injected by GX~301-2 into the gas structure is $M_{\rm tot} \approx 0.19$~M$_\odot$.

The data we received from \citet{fortin_constraints_2022} showed that \mbox{GX~301-2} has a peculiar velocity that is orientated southeast, which means that its birthplace is located northwest of its current position. This means that the interaction with the arm-like structure has varied with time. With previous assumptions about the shape of the structure and the parameter values, we now have a distance between GX~301-2 and the gas shell that varies with time. Hence, the solid angle becomes a function of time, and the fraction of stellar winds interacting with the ISM varies as well.

To simplify the calculations, we considered that GX~301-2 entered three phases that the geometry of the problem seems to justify. First, the approaching phase lasts from first SN event to the gas shell. A fraction of stellar winds is injected in a time-dependent way: $x_{\rm frac}(t)$. It is obvious that $x_{\rm frac}$ increases with time: as the HMXB approaches the structure, its interaction cross section increases. Second is the crossing phase : the entire stellar wind is injected in the arm. It lasts from $t_{\rm in}$ to $t_{\rm out.}$. The third and last phase starts when the gas shell is left to the current location of the HMXB. $x_{\rm frac}$ is again time dependent and decreases with time.

When it is distant from $r$, an area $\mathcal{A}$ covers a fraction of the sphere of radius $r$,
\begin{equation}\label{x_frac}
    x_{\rm frac}(r)=\frac{\mathcal{A}}{4\pi r^2}
.\end{equation}

Here, $\mathcal{A}$ corresponds to the area of the arm-like structure as seen from GX~301-2, and $r$ is the distance between the arm-like structure and the HMXB.
Hence, the total mass injected in the ISM during the first phase is given by
\begin{equation}\label{mass}
    M = \int_0^{t(x_{\rm frac}=1)}\dot{M}x(r)dt 
,\end{equation}
with $t_(x_{\rm frac}=1)$ the time at which we consider that phase 2 began.
Because $r$ decreases with time, we can write
\begin{equation}\label{radius}
    r=r_{\rm init}-vt
,\end{equation}
where $r_{\rm init}$ is the initial distance of GX~301-2 from the gas shell at the first SN event location.

Injecting Eq. (\ref{radius}) into Eq. (\ref{x_frac}), then computing Eq. (\ref{mass}), we obtain  the final result:
\begin{equation}
    M_1=\frac{\dot{M}\mathcal{A}}{4\pi v} \left (\sqrt{\frac{4 i}{\mathcal{A}}}-\frac{1}{r_{\rm init}}\right ) = 0.80^{+0.05}_{-0.03} ~ 10^{-2} M_\odot
.\end{equation}

For phase 3, the calculations are driven in the same way, but we replace Eq. (\ref{radius}) by 
\begin{equation*}\tag{3. bis}
    r = r_{\rm min} + vt
,\end{equation*}
and the  integration interval by $[0,t_{\rm f}]$ , with $r_{\rm min}$ being such that $x(r_{\rm min})=1$ and $t_{\rm f}$ such as $r(t_{\rm f}) = d = 1.3 $~pc the current distance between GX~301-2 and the ISM.

This leads to the final mass formula,\begin{equation*}\tag{4. bis}
    M_3=\frac{\dot{M}\mathcal{A}}{4\pi v} \left (\sqrt{\frac{4\pi}{\mathcal{A}}} -\frac{1}{d}\right ) = 0.65 ~ 10^{-2}M_\odot
.\end{equation*}

For the second phase, we consider that GX~301-2 crosses the arm. This implies that all the stellar wind is injected during this phase, as we assume that interaction occurs for the entire stellar wind reaching the arm-like structure. The crossing time is
\begin{equation}
    t_{\rm cross} = \frac{a}{v}
,\end{equation}
with $a$ being the thickness of the shell, which is estimated to be~0.5~pc using the HAWK-I and Herschel image, and assuming GX~301-2 crossed the shell along the plan of the images. The total injected mass is then

\begin{equation*}\tag{4. ter}
    M_2=\dot{M}\times t_{\rm cross} = 1.60 ~ 10^{-2}M_\odot
.\end{equation*}

Calculations give $M_{\rm tot}  = 3.05^{+0.05}_{-0.03} ~ 10^{-2}$~M$_\odot$ 
, with $M_{\rm tot}$ being the total mass injected during each of the three phases described earlier.

However, injected matter might not always interact with the ISM.
We assumed a simple convective approach, and considered that the winds interact when the kinematic energy density of the wind particles equals the thermodynamic energy of the ISM, that is to say,

\begin{equation}
\frac{1}{2}N_{\rm w}mv_{\rm w}^2 = \frac{3}{2}N_{\rm ISM}k_{\rm B}T_{\rm ISM}
,\end{equation}
$N_{\rm w}$ being distance dependent because of the proper motion of GX~301-2. Hence, we can find an upper distance limit above which the kinetic energy of the stellar wind becomes lower than the thermodynamic energy of the ISM. We retrieved density and temperature parameters for a cold neutral medium from \citet{ferriere_interstellar_2001}.

This gives us the distance
\begin{equation}
    d_{\rm crit} = \left (\frac{N_{\rm w}\mathcal{A}^2mv_{\rm w}^2}{4N_{\rm ISM}k_{\rm B}T_{\rm ISM}} \right )^{1/4}
.\end{equation}

We obtained~$d_{\rm crit} \approx 470$~pc, which means the entire stellar wind that reaches the gas shell can indeed interact.

This implies that the stellar wind had a sufficient energy at all epochs in the past to interact with the ISM. Considering $M_{\rm tot}$ as the mass injected into the ISM that can interact, and a wind velocity of~305~km~s$^{-1}$, we obtain a power of~$\approx 10^{33}$~erg~s$^{-1}$ injected into the ISM on average. This can be used for a comparison with typical stellar jets whose high-energy flux is known to create strong interaction with the ISM. For instance, Cyg~X-1 jet has a power flux of a few~$10^{37}$~erg~s$^{-1}$ for a total energy deposited of~$\approx 7 \times 10^{48}$~erg, as derived by \citet{gallo_dark_2005}. For \mbox{GX~301-2}, in a~500~kyr lifespan, a total energy of a few~$10^{46}$~erg has been deposited in the arm, therefore, approximately~350~times less energy than for a jet. We can also compare it to the W50-SS~433 jet source, for which \citet{margon_rapid_1984} proved a power of a few times~$10^{39}$~erg~s$^{-1}$ and \citet{dubner_high-resolution_1998} retrieved a total energy injected in the ISM of~$10^{51}$~erg in total.
However, jets and winds have different behaviors, and these differences should be addressed.
\citet{miller-jones_searching_2008} argued the need  for a low peculiar velocity of an HMXB to account for the inflation of interstellar bubbles in the case of jet ejection, due to the unidimensional direction. On the other hand, stellar winds are almost isotropic, with no favored direction. Hence, a high peculiar velocity should not prevent the expansion of the interstellar bubble, although the expansion rate may be asymmetric due to a change in the interaction cross section.

Hence, we can deduce that the energy injected by stellar winds of GX~301-2 in the ISM is low, and might not be sufficient to trigger stellar formation, compared to the energy of some powerful jets in our galaxy. We can also add the search for interactions between the microquasar GRS~1915+105 and the ISM \citep{chaty_search_2001}. The microquasar emitted in a luminosity range of~$10^{37}-10^{39}$~erg~s$^{-1}$. The observations for this study were also inconclusive, despite a higher apparent luminosity. This is corroboration for our case here.

In a comparative study of ionizing radiation and stellar wind momentum transfer in the ISM, \citet{haid_relative_2018} showed that for a cold neutral medium, momentum transfer should be dominated by ionizing radiation. However, the order of magnitude of the transferred momentum does not change radically (see their Figure 7), so that the orders of magnitude we computed remain valid.

Finally, we can compute the limiting flux in which the HAWK-I imager can detect a signal. With a derived limiting magnitude of 20.56, we have a limiting luminosity of $\approx 1.79\times 10^{20}$~W in the $H_2$ filter (which is the most sensitive of the four). Considering the total region in which the arm is located, we conclude that this particular region did not emit more than~$\approx 10^{36}$~erg during the exposure time of the instrument.
This would mean that only a very small fraction of the energy coming from the stellar wind of GX~301-2 and absorbed by the arm is indeed processed and emitted by the arm in the form of infrared radiation.

\section{Discussion and conclusions}

Our analysis of HAWK-I images did not allow us toy conclude about the interaction of GX~301-2 with its environment because we did not detect the ISM shell as expected: the SNR between the ISM and the sky background is too low.
The idea of using HAWK-I comes from the wide-field capabilities of the instrument ($7.5'\times 7.5'$), and from its narrow and wide filters. The arm is less bright in the detection range of HAWK-I than for \textit{Herschel} or \textit{Spitzer}, however. The sensitivity of the detector would have required an extended observation time of 2 hours per narrow filter and 1h30 per broadband filter to detect the arm because it is $\text{about three}$  times less luminous at 4.5 $\si{\micro\meter}$ than the dust clump that is detected. We retrieved useful data for the shape of the ISM shell using \textit{Herschel} images, which we modeled as a cylinder to facilitate the calculations. This model could be improved by changing the geometry of the problem to something more realistic, but this would make the numerical calculation more complex. To compute the overall mass transfer, we used the Gaia EDR3 survey to compute the peculiar velocity of GX~301-2, and hence calculated the cross section of the stellar wind interaction with the arm. One of the limitations of the calculations is the hypothesis that the binary system has moved orthogonally to a plane of symmetry of the cylinder. However, it is more likely that it approached it with a certain inclination angle $i$ that has not been considered in the study. The geometry of the system might influence the results derived in this article. Moreover, uncertainties with respect to the age of the first SN event are the main source of uncertainty. We used a fixed SN age of~500~kyr in the past (\citet{simaz_origin_2022}) that was inferred from detailed binary evolution simulations, but uncertainties remain about this parameter. We located the SN event at $500^{+500} _{-250}$~kyr, and took raw estimates for the error bars. This is probably the largest source of uncertainty, but it affects the variation in energy that is transferred to the arm only little, adding an uncertainty of less than 10\%. Hence, the overall conclusion is that winds from GX~301-2 do not seem to have affect the ISM. Another strengthening argument lies in the energetic comparison with the energy of power jets with the ISM, which also seems to unveil that stellar winds are not likely to strongly influence, and even less to trigger, stellar formation by injecting material in the ISM.
Jets have a stronger energy output, but in a narrower area. Their geometry is therefore more efficient in inflating ISM bubbles. In this case, GX~301--2 is am HMXB with a high peculiar velocity of $v_{\rm pec} =56.3^{+3.4}_{-3.2} \,\mathrm{km\,s}^{-1}$ (see \citeauthor{fortin_constraints_2022}, \citeyear{fortin_constraints_2022}), which implies that it could not have stayed long enough in the vicinity to have a significant impact on the ISM. HMXBs with low $v_{\rm pec}$ and stronger stellar winds could inject enough material to inflate ISM bubbles in a way similar to jets, despite their different geometry. In this case, we would not expect two symmetrical lobes from the binary, but rather a higher-density bubble that surrounds the binary.
We note that in our study, the most important phase was the crossing period $t_{\rm cross}$, when half of the total transferred matter was injected in a very short period of time. A binary with a lower $v_{\rm pec}$ could change the total amount of material injected in the ISM due to stellar winds, up to a certain limit.
The characteristics of the ISM also play a role, as shown by \citet{haid_relative_2018}, since the density and temperature of the ISM will affect the way in whichmomentum is transferred to the ISM. This causes an inversion of the ratio of stellar winds and ionizing radiation contributions for a warm neutral medium.

Multiwavelength observations of jets interacting with the ISM have been proven successful in helping us understand the mechanisms in place. For instance, the comprehensive work of \citet{chaty_search_2001} on GRS~1915+105, which reported observations at various wavelengths, have helped to constrain the impact of jets on the ISM. Following previous studies on SS 433, the authors concluded that molecular clumps can be common where highly energetic sources interact with the ISM, and can be detected with high-density tracers. The present study is part of a similar multiwavelength analysis of the interaction between stellar winds and the ISM. Following the work initiated by \citet{servillat_herschel_2014} in the IR domain with \textit{Herschel} , but also the modeling of its SED ranging from 0.4 to $4 \times 10^4 \, \si{\micro\meter}$, we studied the J and K bands of the spectrum in detail, constraining the interaction among the stellar winds and the ISM. 
This might be extended to the millimeter and submillimeter regions, however, where we expect that research in the millimeter spectrum at around 100GHz could be of interest to detect tracers of high density, such as HCO+ or CO (see \citet{chaty_search_2001} ). Overall, \citeauthor{tetarenko_mapping_2018} (\citeyear{tetarenko_mapping_2018}, \citeyear{tetarenko_jet-ism_2020}) showed in a series of articles that molecular line studies in millimeter and submillimeter bands provide information about the interaction between XRBs and the ISM, mostly through jets. A search for similar interactions with stellar winds might provide new insights into the interaction between stellar winds and the ISM.
In contrast to jets, strong shocks are not expected from from stellar wind that interacts with the ISM medium. Therefore, radio maps would probably not be best suited for studying these interactions, as concluded by \citet{kaiser_revision_2004}.
In addition, a 20 cm line analysis could also show an ionization front, similarly to what \citet*{rodriguez_large_1998} revealed for GRS 1915+105.
As mentioned previously, the characteristics of the ISM might influence the impact of stellar wind on the ISM. New studies in these wavelengths might therefore help to constrain these parameters and allow us to understand the feedback of GX 301-2 better, which is  an HMXB with strong stellar winds. It would also allow us to understand its surrounding environment better.

We used IR images from different instruments and surveys to investigate the impact of GX~301-2 on the surrounding ISM shell. We reduced data from the HAWK-I instrument and \textit{Herschel}/PACS instrument, and used them to detect the arm structure for the remaining analysis. In order to answer the question of the interaction between GX~301-2 and the shell through stellar winds, we computed an estimate of the total mass lost by GX~301-2 and injected in the arm. Different models were considered, from a fixed source to a moving source. We used simple energy considerations with a convective model to verify whether energetic contribution of the stellar wind to the arm was relevant. We then computed the total energy injected by the stellar winds of GX~301-2 in the ISM, and compared it to some known jets that are produced by galactic binaries whose interaction with the ISM has previously been proven by different studies.
Finally, we concluded that stellar wind from GX~301-2 is unlikely to have fed the ISM enough energy to trigger stellar formation in this particular region.

\begin{acknowledgements}
      This research has made use of Aladin sky atlas and the SIMBAD database developed and maintained at CDS, Observatoire de Strasbourg;  Astropy, a community-developed core Python package for Astronomy (Astropy Collaboration, 2013); IRAF, distributed by the National Optical Astronomy Observatory, operated by the Association of Universities for Research in Astronomy (AURA) under a cooperative agreement with the National Science Foundation; and NASA's Astrophysics Data System bibliographic Services. SC acknowledges the CNES (Centre National d’Études Spatiales) for the funding of MINE (Multi-wavelength INTEGRAL Network). The authors are grateful to the LabEx UnivEarthS for the funding of Interface project I10 «From binary evolution towards merging of compact objects».
\end{acknowledgements}

%
%
\bibliographystyle{aa}
\bibliography{Bibliographie-GX-301-2}

\end{document}